\documentclass[10pt,letterpaper,aps,pra,twocolumn,superscriptaddress]{revtex4-1}
\usepackage{amsmath}
\usepackage{amsfonts}
\usepackage{amssymb}
\usepackage{bm}
\usepackage[pdftex]{graphicx}
\usepackage{epstopdf}
\usepackage{etoolbox}
\usepackage{times}
\usepackage[letterpaper,margin=1in]{geometry}
\usepackage[utf8]{inputenc}
\usepackage{color}
\usepackage[usenames,dvipsnames]{xcolor}
\usepackage{epstopdf}
\usepackage{units}
\usepackage{natbib}
\usepackage[normalem]{ulem}
\usepackage{enumerate}
\usepackage[caption=false]{subfig}
\usepackage{longtable}

\newcommand{\rem}[1]{{\color{red}\sout{#1}}}

\usepackage{float}

\begin{document}
\title{Weak-value amplification and optimal parameter estimation in the presence of correlated noise }

\author{Josiah Sinclair}
\affiliation{Centre for Quantum Information and Quantum Control, and Institute for Optical Sciences, Department of Physics,  University of Toronto, 60 St George Street, Toronto, Ontario, Canada M5S 1A7}
\author{Matin Hallaji}
\affiliation{Centre for Quantum Information and Quantum Control, and Institute for Optical Sciences, Department of Physics,  University of Toronto, 60 St George Street, Toronto, Ontario, Canada M5S 1A7}
\author{Aephraim M.  Steinberg}
\affiliation{Centre for Quantum Information and Quantum Control,
and Institute for Optical Sciences, Department of Physics,
University of Toronto, 60 St George Street, Toronto, Ontario, Canada M5S 1A7}
\author{Jeff Tollaksen}
\affiliation{Institute for Quantum Studies, Chapman University, 1 University Drive, Orange, CA 92866, USA}
\affiliation{Schmid College for Science and Technology, Chapman University, 1 University Drive, Orange, CA 92866, USA}
\author{Andrew N. Jordan}
\affiliation{Department of Physics and Astronomy \& Center for Coherence and Quantum Optics, University of Rochester, Rochester, New York 14627, USA}
\affiliation{Institute for Quantum Studies, Chapman University, 1 University Drive, Orange, CA 92866, USA}

\date{\today}
\begin{abstract}
We analytically and numerically investigate the performance of weak-value amplification (WVA) and related parameter estimation methods in the presence of temporally correlated noise. WVA is a special instance of a general measurement strategy that involves sorting  data into separate subsets based on the outcome of a second ``partitioning'' measurement. Using a simplified noise model that can be analyzed exactly together with optimal statistical estimators, we compare WVA to a conventional measurement method. We find that introducing WVA indeed yields a much lower variance of the parameter of interest than does the conventional technique, optimized in the absence of any partitioning measurements. In contrast, a statistically optimal analysis that employs partitioning measurements, incorporating all partitioned results and their known correlations, is found to yield an improvement -- typically slight -- over the noise reduction achieved by WVA. This is because the simple WVA technique is not tailored to a given noise environment and therefore does not make use of correlations between the different partitions. We also compare WVA to traditional background subtraction, a familiar technique where measurement outcomes are partitioned to eliminate unknown offsets or errors in calibration. Surprisingly, in our model background subtraction turns out to be a special case of the optimal partitioning approach in the balanced case, possessing a similar typically slight advantage over WVA. These results give deeper insight into the role of partitioning measurements, with or without post-selection, in enhancing measurement precision, which some have found puzzling. They also resolve previously made conflicting claims about the usefulness of weak value amplification to precision measurement. We finish by presenting numerical results to model a more realistic laboratory situation of time-decaying correlations, showing our conclusions hold for a wide range of statistical models.
\end{abstract}

\newcommand{\op}[1]{\hat{\bm #1}}                
\newcommand{\ket}[1]{\lvert#1\rangle}
\newcommand{\bra}[1]{\langle#1\rvert}
\newcommand{\pr}[1]{\ket{#1}\bra{#1}}
\newcommand{\ipr}[2]{\langle #1 | #2 \rangle}
\newcommand{\mean}[1]{\left\langle #1 \right\rangle}
\newcommand{\cw}{\circlearrowright}
\newcommand{\ccw}{\circlearrowleft}
\newcommand{\be}{\begin{equation}}
\newcommand{\ee}{\end{equation}}
\newcommand{\bea}{\begin{eqnarray}}
\newcommand{\eea}{\end{eqnarray}}
\newcommand{\ra}{\rangle}
\newcommand{\la}{\langle}

\maketitle

\section{Introduction}

Weak-value amplification (WVA) \cite{AAV} is a technique that has been used in a variety of experimental settings to permit the precise measurement of small parameters \cite{Hosten2008,Dixon2009,Starling2009,Starling2010,Starling2010b,Hogan2011,turner2011picoradian,pfeifer2011weak,zhou2012experimental,egan2012weak,xu2013phase,viza2013weak,shomroni2013demonstration,qiu2014determination,jayaswal2014observation,magana2014amplification,salazar2014measurement,zhou2014observation,Viza2014,salazar2015demonstration,salazar2015enhancement}.  Whether or not WVA has an actual advantage in terms of the resulting measurement precision has been the subject of an ongoing debate over the past few years \cite{tanaka2013information, ferrie2014weak, knee2014amplification,combes2014quantum, jordan,Howell, alves2014weak,Pang2014}.  In this paper, we delineate the situations under which WVA does in fact improve measurement precision in the presence of correlated noise, comparing it with competing approaches.  This investigation leads to interesting connections between WVA and other techniques such as background subtraction and lock-in amplification, which elucidates the technical advantages that have already been experimentally observed.

Alongside arguments regarding the usefulness of WVA to precision measurement, there has continued to be a great deal of work extending and improving the basic technique of WVA in other situations.
Some recent advances in the field include the incorporation of photon recycling of discarded events \cite{Dressel2013,lyons2015power}; the observation that WVA can improve measurement precision in cases with detector saturation \cite{jeremieharris2016}, the optimization of the shape of the meter probe \cite{susa2012optimal}, and a generalized approach to probabilistic quantum metrology \cite{calsamiglia2014probabilistic}. Of particular interest is
weak-value amplification with entanglement \cite{Pang2014,pang2014improving},squeezing \cite{pang2015improving}, and the observation that weak-value amplification can suppress systematic errors \cite{PhysRevA.94.012329}, which is closely related to the present work. We refer the reader to recent reviews for a wider overview of this field \cite{Kofman2011,Dressel2014}.

Weak-value amplification involves two measurements. In the first, a known system observable is measured via a weak interaction with a measurement apparatus. The effect of this weak interaction is to induce a small shift in the pointer of the measurement apparatus. The size of this shift is determined by the observable (which is typically known) and the coupling strength, which we are interested in estimating. In weak-value amplification this coupling strength is usually very weak, so that very little information is gained about the state being measured, and the corresponding measurement disturbance is minimized. The second measurement is a strong projective measurement in a different basis on the system. Because the system and measurement device are left weakly entangled by the first measurement, there are interesting correlations between the two systems. These correlations can be seen by dividing the dataset into partitions corresponding to the results of the second projective measurement, and then averaging the different partitions with different weights. In the simplest version of this, the weights are zero and one. This corresponds to discarding certain measurement outcomes based off the result of the second projective measurement - this is called postselection.  
The mean shift of the pointer conditioned upon the post-selection succeeding is called the weak value. It is defined in terms of the initial state of the system ($\ket{i}$), the state the system is found to be in if the postselection succeeds ($\ket{f}$), and the operator associated with the first "weak" measurement, $\bf A$, 
\be
A_w \equiv \frac{\la f | {\bf A} | i\ra}{\la f |  i\ra}.
\ee
The weak value can become quite large when the overlap of the initial and final state becomes very small. There is a corresponding reduction in the size of the data set which shrinks with the probability of the post-selection succeeding (which is just the square of the overlap of the initial and final states).

 Despite the many articles on the topic, a general theory of when the technique yields a quantitative advantage has so far remained elusive. It is easy to see that WVA yields no advantage in precision for optical experiments that are shot-noise--limited \cite{Starling2009}, taking the number of input photons as the resource. Nevertheless, experimental metrological works such as Refs.~\cite{Hosten2008,Dixon2009}, demonstrated that WVA may offer an advantage in the presence of technical noise sources. In 2011, Feizpour, Xing, and Steinberg addressed the case of additive, time-correlated, Gaussian random noise and argued that weak-value amplification can result in an improvement in the signal-to-noise ratio of the measured parameter \cite{feizpour2011amplifying}. More recently, others have argued the WVA method to be inherently sub-optimal because it involves discarding a portion of the measurement outcomes \cite{ferrie2014weak}. Defenders of WVA have responded that this comparison was to a theoretically optimal but experimentally challenging approach, whereas previous analyses had compared WVA to conventional techniques \cite{LevComment}. Also, it has been shown that in the ideal (uncorrelated noise case) weak-value measurement the information contained in the discarded measurement outcomes is a tiny fraction of the total information, despite the discarded outcomes making up the vast majority of the total number of outcomes, and that the technique is therefore asymptotically optimal \cite{jordan,Howell,alves2014weak,Pang2014}. We will revisit the case of correlated noise and show that while WVA is much better than the conventional approach, in the slow noise case it is (typically) slightly inferior to a statistical analysis which optimally utilizes all partitioned outcomes and the correlations between them. We will also show that WVA and the optimal partitioning approach are closely related to background subtraction and lock-in amplification techniques which are well known. 

This paper is organized as follows.   In Sec.\ref{tools}, we introduce the tools used in estimation theory: estimators; variance; the Cram\'er-Rao Bound; our metric of choice, the Fisher Information; and a model for correlated Gaussian noise. In Sec. \ref{2dexample}, we explore the Fisher Information as an information metric using a simple two-measurement-outcome example.  In Sec.\ref{evalueanalysis}, we give an eigenvalue analysis of the Fisher information, and show it may be expressed as a weighted average of the eigenvalues of the covariance matrix.  The weak-value amplification effect is introduced in Sec.\ref{wva}. An optimal partitioning measurement approach is introduced in Sec.\ref{improved}, which improves slightly on the advantage achieved by WVA over the direct method by including all partitioning states and the correlations between them.  This physics is illustrated in Sec.\ref{solvable}, where an exactly solvable model is introduced, and the variance of all the estimation strategies is given explicitly and compared.  A numerical investigation  of these issues is presented in Sec.\ref{numerics}, where an experimentally motivated correlated noise model is given, and analyzed.  Our conclusions are given in Sec.~\ref{conclusions}.

\section{Parameter Estimation}
\label{tools}

Consider a common scenario in the natural sciences where the goal is the measurement of some unknown parameter. We represent our data set $\{s_i\}$, where $i$ identifies the $i$th measurement, $i = 1, 2, ... , N$. Measurement outcomes can be scaled and shifted so the measured signal $s_i$ can be modeled as the parameter of interest, $d$, and noise, $x_i$, 
\be
s_i = d+x_i
\ee
where $x_i$ are zero mean Gaussian random variables \cite{feizpour2011amplifying}. In the most generic case, these variables may be correlated, and we define the correlation function  $C_{i,j}(x) = \la x_i x_j \ra$ - this is also called the covariance matrix. Our task is to estimate the unknown parameter $d$, from the data set and our knowledge of the covariance matrix. In this paper, we consider the covariance matrix known, and the detailed knowledge able to be applied to implement the optimal estimator of the unknown quantity $d$.  The extra resources required to estimate this matrix and implement the optimal estimator are not considered \cite{jordan}. By choice of estimator, we simply mean some algorithm which maps our data set to an estimate of $d$. Taking the arithmetic average corresponds to the most conventional and straightforward estimator. Because the noise has zero mean, this estimator is unbiased, which means that in the limit of $N\rightarrow \infty$, the estimate of the parameter converges to $d$. There are many other unbiased estimators which could be constructed depending on how the measurement is designed and implemented.

We will discuss two different classes of measurement design. The first class, which most conventional measurements fall into, we term ``direct". ``Direct'' measurements only involve measurements on the parameter of interest.  The second class, which we term ``partitioning", introduces a second measurement that is used to sort the first measurement's outcomes into different partitions. Often, ``partitioning" class measurements possess some advantage versus ``direct" measurements, because they can exploit correlations between different partitions. The simplest ``partitioning" class measurement involves discarding all data points which fail to meet some criterion assessed by the second measurement. This is post-selection. While others have others have pointed out that instead of discarding outcomes, they should be weighted and optimally analyzed, our assertion is just that exploiting such correlations may have significant advantages, and that while the actual throwing out of data per se can never increase the amount of information available, there may be regimes where some (perhaps most) of the advantage survives even this procrustean approach. 
\begin{figure}[H]
\includegraphics[width=\columnwidth]{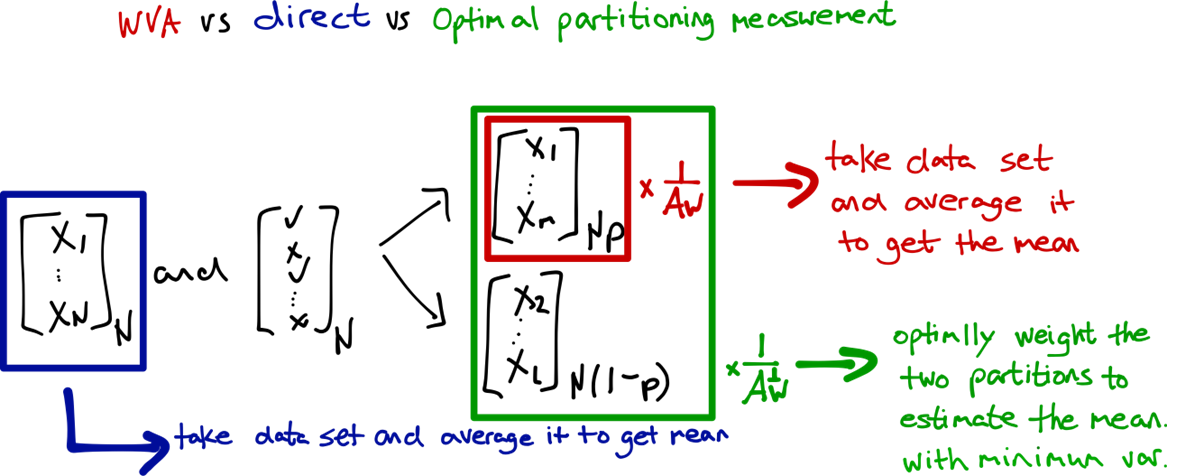}
\caption{Here we schematically represent the three different measurements we are considering, ``direct", involving measurements on the parameter of interest, WVA, which involves a post-selected set of data, and the ``optimal partitioning measurement", where both post-selected and post-selected-rejected partitions are retained and optimally analyzed.
}
\label{visualdepict}
\end{figure}
Once the ``measurement design" has been specified, the space of available estimators is infinite. As the variance associated with different estimators can vary wildly, the ``choice of estimator" is highly nontrivial. Conveniently, a technique called maximum likelihood estimation \cite{jordan} can be used to identify the estimator with minimum variance in the large dataset limit. Furthermore, there is a mathematical theorem \cite{kay2010fundamentals} that bounds the minimum variance of an unbiased estimator to be larger than the inverse \textit{Fisher Information}. This allows us to forgo discussion of estimators entirely, skipping directly to the calculation of the Fisher Information, which is defined as the inverse variance of the optimal estimator. Find the Fisher Information for a given measurement scenario, invert it, and one will have found the minimum variance of any unbiased estimator.

\section{The Fisher Information in a two measurement outcome example}
\label{2dexample}

It is instructive to consider a pedagogical example that will clarify the meaning of the Fisher Information,  the importance of the choice of estimator, and the significance of the covariance matrix to this task.  If we return to the scenario in the previous section, but imagine collecting only two data-points, we will have a data set ($s_1, s_2$) with assumed knowledge of the covariances and variances. As before our goal is to estimate the mean $d$.  The covariance matrix, which is a 2x2 matrix is:
\be
C(s) = \begin{pmatrix} {\rm Var}(s_1) &  {\rm Cov}(s_1, s_2) \\{\rm Cov}(s_2, s_1)  & {\rm Var}(s_2)
\end{pmatrix},
\ee
where ${\rm Var}(s_1),  {\rm Var}(s_2) \ge 0$, and ${\rm Cov}(s_1, s_2) = {\rm Cov}(s_2, s_1)$ are all real numbers. The Cauchy-Schwartz inequality (${\rm Cov}(s_1, s_2)^2 < {\rm Var}(s_1)  {\rm Var}(s_2)$) guarantees both eigenvalues be non-negative which also guarantees $C$ to be positive semidefinite. The ${\rm Cov}(s_1, s_2)$ itself can be positive or negative, representing either positive or negative correlation between $s_1$ and $s_2$.  Our goal is to estimate the expectation value of our two data points while minimizing the variance of our estimate. Were we to use the first data point alone to estimate the mean, the variance would be ${\rm Var}(s_1)$; similarly, use of the second data point alone gives $ {\rm Var}(s_2)$.  Instead, we can use a combination of the two data points as our estimator, $s = \alpha s_1 + \beta s_2$, where $\alpha$ and $\beta$ are constant weighting factors.  To keep this estimator unbiased we require that $\alpha + \beta=1$. The variance of our estimator $s$ is given by
\be
{\rm Var}(s) = \alpha^2 {\rm Var}(s_1) + \beta^2  {\rm Var}(s_2) + 2 \alpha \beta {\rm Cov}(s_1, s_2).
\label{vars}
\ee
We minimize the variance with respect to $\alpha$, keeping ${\rm Var}(s_1), {\rm Var}(s_2), {\rm Cov}(s_1, s_2)$ fixed.  Doing so gives a minimum variance of 
\be
{\rm Var}(s)_{min} = \frac{{\rm Var}(s_1)  {\rm Var}(s_2)-{\rm Cov}(s_1, s_2)^2}{{\rm Var}(s_1)+ {\rm Var}(s_2) - 2 {\rm Cov}(s_1, s_2)},
\ee
which is smaller than either ${\rm Var}(s_1)$ or $ {\rm Var}(s_2)$ for any allowed values of ${\rm Var}(s_1),  {\rm Var}(s_2), {\rm Cov}(s_1, s_2)$.  We note that unlike uncorrelated random variables, the inverse variance is not additive. If instead we took equal weighting of the two data points, $\alpha=\beta=1/2$, this would give 
\be
{\rm Var}(s)_{equal} = {\rm Cov}(s_1, s_2)/2 + ({\rm Var}(s_1)+ {\rm Var}(s_2))/4.  
\ee
If the two outcomes are perfectly negatively correlated [${\rm Cov}(s_1, s_2) \rightarrow - \sqrt{{\rm Var}(s_1)   {\rm Var}(s_2)}$], then both the optimal and the equal weighting estimators have zero variance. This can be understood as resulting from anticorrelated fluctuations canceling each other out, for example if $s_1 = d + x_1$, and $s_2 = d - x_1$, a straightforward averaging of $s_1, s_2$ will result in the perfect cancellation of $x_1$.  If the two outcomes are perfectly positively correlated, but the variances are not equal (${\rm Var}(s_1) \neq {\rm Var}(s_2)$), the optimal variance vanishes, whereas the equal weighting variance limits to that of using just a single outcome. This can be understood by considering the case where $s_1 = d + 2x_1$, and $s_2 = d + x_1$. If our estimator is $s = 2s_2 - s_1$, we can eliminate the noise just like in the anti-correlated case. When the variances are equal and the correlations are positive the optimal variance does not vanish, because the noise cannot be canceled without making the estimator biased. In this case the equal-weighting estimator turns out to be optimal. We recall that the Fisher Information is equal to the inverse of the variance of the optimal estimator in the large data limit. The Fisher Information is defined for smooth distributions as \cite{kay2010fundamentals}
\be
{\cal I}  = - \int ds_1\ldots ds_n P(s_1, s_2, ... s_n|d) \frac{\partial^2}{\partial d^2} \ln{P},
\label{fidef}
\ee
where $P=P(s_1, s_2, ... s_n|d)$ is the probability distribution of $\{s_1, ... s_n\}$ given a fixed value of $d$. In our model, it is taken as a multi-dimensional Gaussian distribution with mean $d {\bf 1}$ and covariance matrix $C_{ij}$.  From Ref.~\cite{jordan} the Fisher information about the mean for Gaussian correlated noise is given by
\be
{\cal I}  = \sum_{i,j}^{N} [C^{-1}]_{i,j},
\label{fi}
\ee
and the Fisher information for $N=2$ is simply the minimized variance we found previously
\be
{\cal I} = \frac{{\rm Var}(s_1)+ {\rm Var}(s_2) - 2 {\rm Cov}(s_1, s_2)}{{\rm Var}(s_1)  {\rm Var}(s_2) - {\rm Cov}(s_1, s_2)^2} = \rm Var(s)_{min}^{-1}.
\label{fi2by2}
\ee

\begin{figure}[H]
\includegraphics[width=\columnwidth]{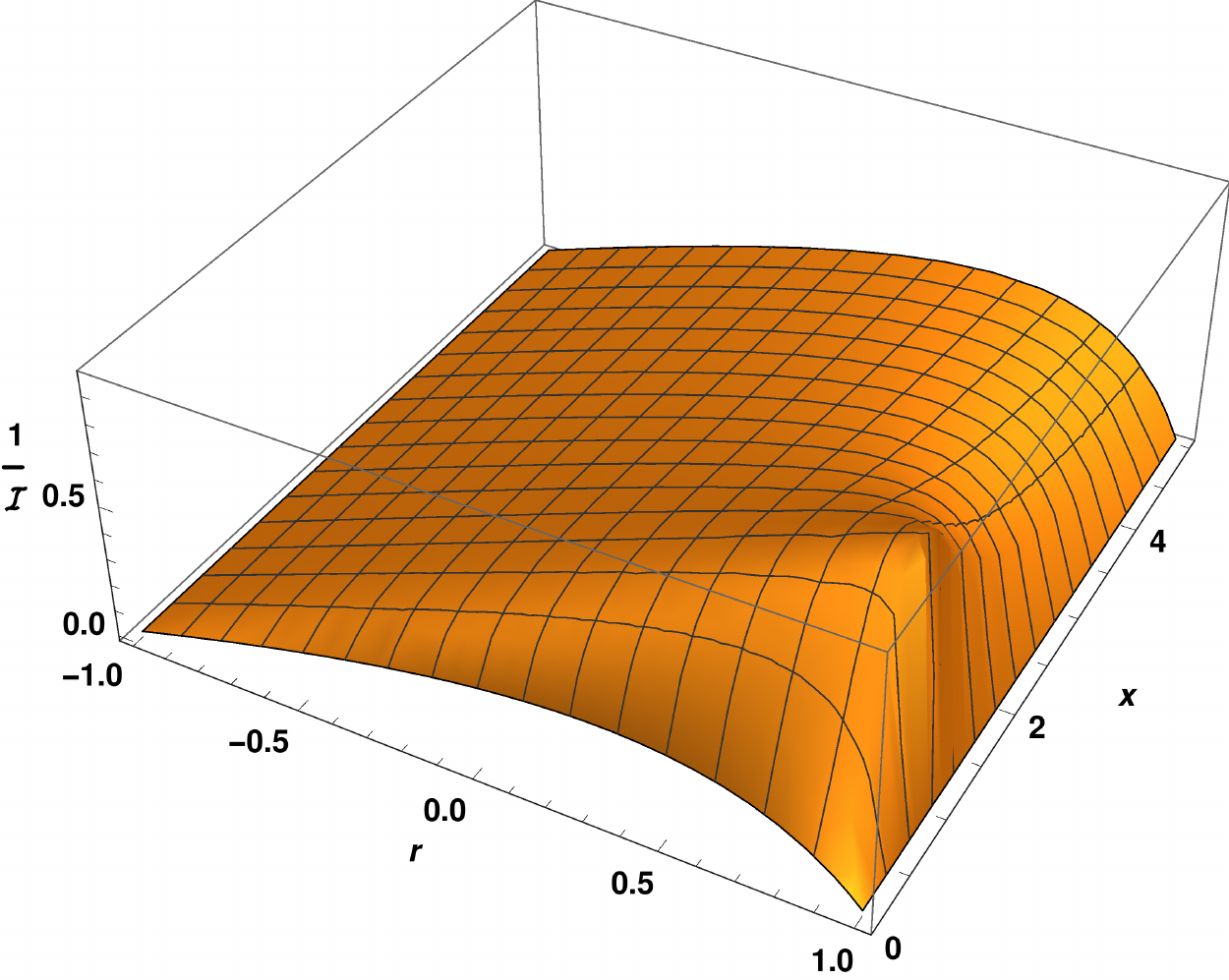}
\caption{Inverse Fisher information (minimum variance of $s$) versus $x$ and $r$, Eq.~(\ref{fi2by2}), plotted in units of $\sqrt{{\rm Var} (s_1) {\rm Var} (s_2)}$.
}
\label{fisher2by2fig}
\end{figure}

It is convenient to introduce two parameters, an asymmetry parameter, $x = {\rm Var}(s_1)/ {\rm Var}(s_2) \in [0, \infty)$, and a relative correlation parameter, $r = {\rm Cov}(s_1, s_2)/\sqrt{{\rm Var}(s_1) {\rm Var}(s_2)} \in [-1, 1]$.  Dividing the inverse Fisher information by $\sqrt{{\rm Var}(s_1) {\rm Var}(s_2)}$ gives a function that depends only on $r$ and $x$. We plot the inverse Fisher information in this case in Fig.~\ref{fisher2by2fig}, noting the asymmetry in both $r$ and $x$ which we will now explore.

\begin{figure}[H]
\includegraphics[width=\columnwidth]{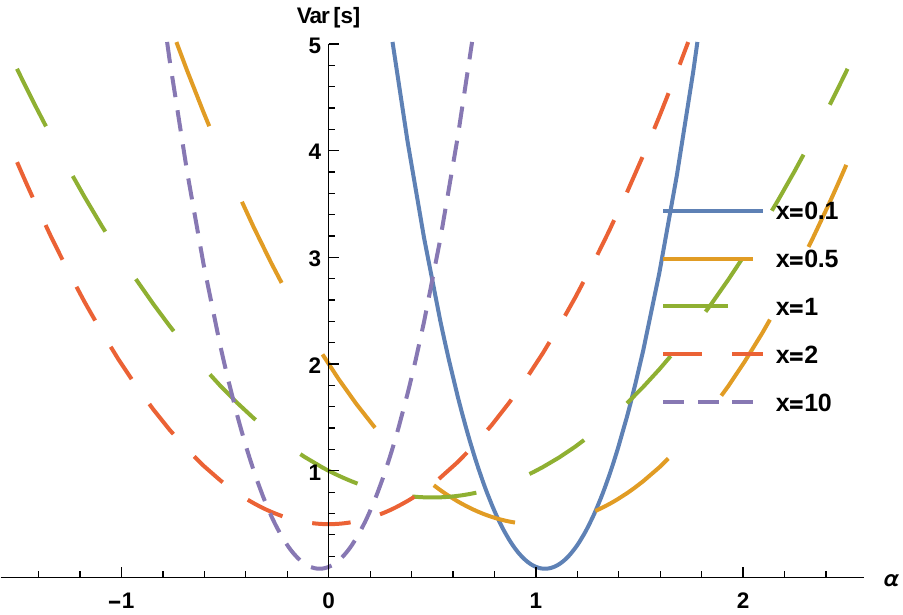}
\caption{Variance of the estimator $s$ versus $\alpha$, Eq.~(\ref{vars}). 
We choose 50\% correlated outcomes ($r=1/2$), and plot for different values of asymetry ($x$).
}
\label{varsfig1}
\end{figure}

\begin{figure}[H]
\includegraphics[width=\columnwidth]{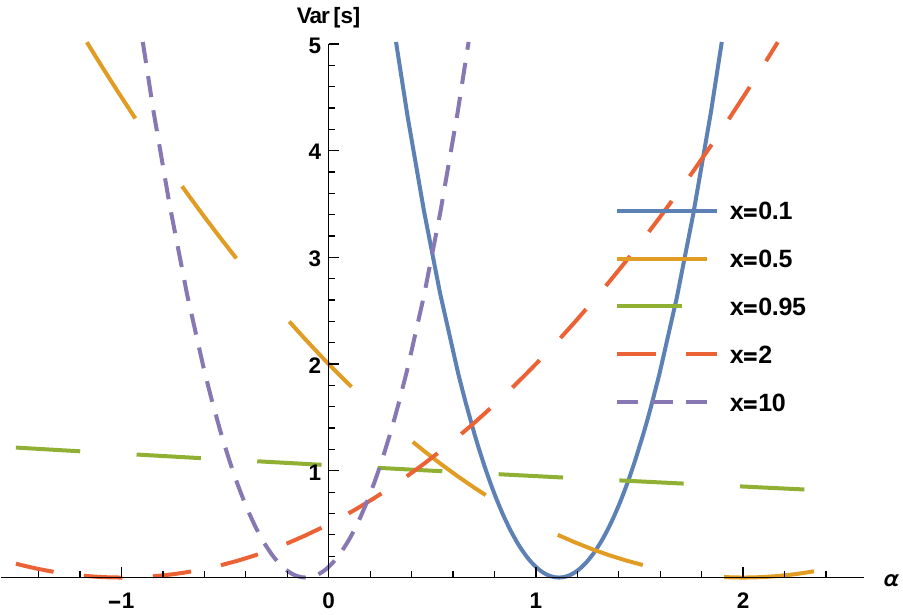}
\caption{Variance of the estimator $s$ versus $\alpha$, Eq.~(\ref{vars}). 
We choose maximally correlated outcomes ($r=1$), and plot for different values of asymmetry ($x$).
}
\label{varsfig2}
\end{figure}

\begin{figure}[H]
\includegraphics[width=\columnwidth]{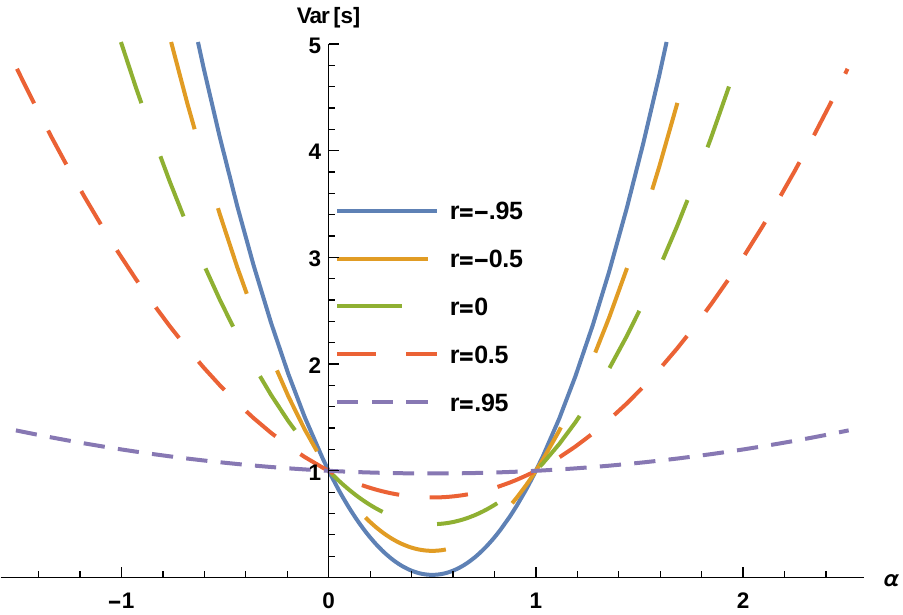}
\caption{Variance of the estimator $s$ versus $\alpha$, Eq.~(\ref{vars}) 
We choose the symmetric case ($x=1$), and plot for different values of correlation between the outcomes ($r$).
}
\label{varsfig3}
\end{figure}

We note that (\ref{fi2by2}) indicates that negative values of ${\rm Cov}(s_1, s_2)$ (or $r$) typically have higher Fisher information than positive values; that is, anti-correlation is more informative than correlation. In  Figs.~\ref{varsfig1},\ref{varsfig2},\ref{varsfig3}, we plot the variance of $s$ versus $\alpha$ for different values of $x$ and $r$.  The minimum value of the estimator corresponds to the inverse Fisher information for that choice of covariance matrix. These figures highlight how the information that can be extracted from a probability distribution depends in a complicated way on the parameters of that distribution even in the simplest (two dimensional) cases. Importantly, they show that while anticorrelations typically increase the available information, positive correlations can in certain circumstances also boost the amount of information that is available. 
In what follows we will show that, due to the properties of the Covariance matrix, the Fisher Information can be expressed simply in terms of the eigenvalues and eigenvectors of $C$. This will pave the way for an exactly solvable noise model which we will use to address the questions raised in section I, by comparing the Fisher Information of the various measurement strategies. Finally, we will present a numerical investigation where the conclusions reached with the exactly solvable noise model are shown to hold in experimentally realistic scenarios.

\section{Eigenvalue Analysis of the Fisher Information}
\label{evalueanalysis}
We recall that the covariance matrix $C$, is a symmetric, positive semidefinite matrix.  Therefore, we can make an orthogonal decomposition of it as follows,
\be
C = O D O^T,
\ee
where $O$ is an orthogonal matrix, $O^T$ is its transpose, and $D$ is a diagonal matrix with eigenvalues $\sigma^2_j > 0$ for all $j$, since it is positive definite.  We do not consider $\sigma_j =0$ for any $j$ since that corresponds to a deterministic outcome, which then gives infinite information (or zero variance).  It is then easy to see that the inverse of $C$ is given by
\be
C^{-1} = O D^{-1} O^T,
\ee
by direct calculation, where $D^{-1}$ is a diagonal matrix with elements $\sigma^{-2}_j>0$ for all $j$. 

Substituting that decomposition to Eq.~(\ref{fi}) gives
\be
{\cal I} = \sum_{i,j} \sum_k O_{i,k} \sigma^{-2}_k O^T_{k,j} = \sum_k \sigma^{-2}_k \sum_i O_{i,k} \sum_j O_{j,k}.
\ee
We define a vector ${\bf v}$ with components, $v_k = \sum_{i} O_{i,k}$.  This then gives
\be
{\cal I} = \sum_k \sigma^{-2}_k v_k^2.
\ee
Next we note that the $O$ matrix is orthogonal, and therefore $O O^T = I$, the identity matrix.  Summing over both indices we obtain
\be
\sum_{i,j,k} O_{i,k} O^T_{k,j} = \sum_k v_k^2 = \sum_{i,j} \delta_{i,k} = N.
\label{normalization}
\ee
Since the sum of the squares of the $v$ vector components must be $N$ by orthogonality, we define a weight vector ${\bf w}$, whose elements are $w_i = v_i^2/N$, so that the sum of the weights is 1.
With this definition, the Fisher information is
\be
{\cal I} = N \sum_k \sigma^{-2}_k w_k.
\label{FI}
\ee
Thus, this Fisher information is given by the weighted average of the eigenvalues of the covariance matrix, with the weights related to the eigenvectors of covariance matrix.

According to the Cram\'er-Rao inequality (CRI), the variance of any unbiased estimator $\hat d$ must be greater than the inverse Fisher information \cite{kay2010fundamentals},
\be
{\rm Var}[{\hat d}] \ge {\cal I}^{-1}.
\ee
Let us see if this is true in the case of the simple estimator ${\hat d} = (1/N) \sum_j s_j$.  The variance is given by
\be
{\rm Var}[{\hat d}] = (1/N^2) \sum_{i,j} C_{i,j}.
\label{direct}
\ee
We can make a similar analysis as above, $C = O D O^T$, to find
\be
{\rm Var}[{\hat d}] =(1/N^2)\sum_{i,j,k} \sigma_k^2 O_{i,k} O_{j,k}.
\ee
We can rewrite this as
\be
{\rm Var}[{\hat d}]= (1/N)\sum_k \sigma^2_k w_k.
\label{weights-var}
\ee
The CRI can be restated in this case as the inequality
\be
\sum_i w_i \sigma_i^2 \sum_j w_j \sigma_j^{-2} \ge 1.
\label{desired}
\ee
This relation can be proved directly with the Cauchy-Schwarz inequality.  We define a vector ${\bf u}_1$ of dimension $N$ with elements $u_{1,i} = \sqrt{w_i} \sigma_i$, and another vector ${\bf u}_2$ of the same dimension, with elements $u_{2,i} = \sqrt{w_i} \sigma_i^{-1}$.  The Cauchy-Schwarz inequality,
\be  
| {\bf u}_1 \cdot {\bf u}_2 | \le ||  {\bf u}_1||  \, || {\bf u}_2 ||,
\label{ti}
\ee 
applied to these vectors gives $\sum_i w_i \sigma_i/\sigma_i= 1$ for the left hand side of (\ref{ti}), since $w_i$ are weighting factors, and  $(\sum_i w_i \sigma_i^2  \sum_j w_j \sigma_j^{-2})^{1/2}$ for the right hand side of (\ref{ti}).  If a square root of a quantity is greater than 1, that quantity is greater than 1 as well, establishing the desired relation (\ref{desired}).

We note that a sufficient condition on $C$ to make the simple estimator efficient is that $C$'s rows (or columns) all sum to the same number.  This is equivalent to the statement that $C$ has an eigenvector ${\bf e}_1 = (1, 1, \ldots, 1)^T$.  By construction, all other eigenvectors are orthogonal to this one, so the vector ${\bf v}$ will have components 0, except for the first entry.  Therefore, the weighting factors are given by $w_j = (1, 0, 0, \ldots 0)$.  This gives the variance (\ref{weights-var}) ${\rm Var}[{\hat d}] = \sigma_1^2/N$, which saturates the CRB, as seen from Eq.~(\ref{FI}).  In this case, $\sigma_1^2$ is just the sum of any row or column of $C$.  As we will see in Sec.~\ref{solvable} this result will facilitate calculating the Fisher Information for our simple correlated noise model.  

\section{Including the Weak-value amplification effect}
\label{wva}
Let us now consider the weak-value case.
In a weak-value-type metrology experiment, a system is weakly coupled to a meter via an interaction whose coupling strength we would like to determine. The interaction term can be written as the product of some system observable, $\textbf{A}$, and an operator on the pointer that we call $\textbf{P}$, multiplied by the coupling strength suggestively labeled $d$. The interaction Hamiltonian is therefore
\be
H_I= d \: \textbf{P} \cdot \textbf{A}.
\ee
The pointer operator is termed $\textbf{P}$, because the effect of $H_I$ is to generate translations $\propto d\: \textbf{A}$ in the conjugate variable, consequently interpreted as a pointer position \cite{Neumann}. For the measurement to be considered ``weak" the pointer shift must be much less than the uncertainty, ($\sigma$), in the position of the pointer, $d \: \la  {\bf A} \ra \leq \sigma$.  
In the standard weak-value approach, the system is pre- and post-selected to be in state $|i\ra$ and $|f\ra$.  For an $N$ dimensional system, there are $N$ simultaneously possible outcomes for a given projective measurement, with the probability of a given outcome $|f\ra$ given by $\gamma = |\la f | i \ra|^2$. While it is possible that in general there may be information in the probability of the selection \cite{Jordan2015,Walsmsley,alves2014weak,huang2015weak}, in the usual approach, \`a la Aharonov, the probability of selection is independent of the parameter of interest, and all of the available information is in the meter deflections \cite{AAV,alves2014weak}.
If the average deflection in the absence of a second selective measurement is given by $d \: \la i | {\bf A}| i \ra = d {\la \bf A \ra } $, then in the weak limit, the Fisher information about $d$ is multiplied by a factor of ${\la \bf A \ra}^2$ (see Eqs.~(\ref{fidef},\ref{fi})).  In contrast,  in the presence of the second selective measurement, the deflection is given by $A_w d$, where
\be
A_w \equiv \frac{\la f | {\bf A} | i\ra}{\la f |  i\ra},
\ee
is defined as the weak value. 
We note that the weak-value can be imaginary and that it is not bounded by the eigen spectrum of the operator $\bf A$. If the probability of post-selection is made very small, the weak-value can become very large, hence the term amplification. In this work, we focus on real weak-values;  see Refs.~\cite{PhysRevA.85.060102,Howell,jordan} for a discussion of imaginary weak-value amplification. 

If we make a weak-value amplification-type experiment, with post-selection of probability $\gamma$, the resulting data set, $\{s_i \}$, contains on average only $\gamma N$ data points (where we recall that $N$ was the number of data points in the non-post-selected measurement). Whereas before we would rescale the meter deflection by the expectation value $\textbf{A}$ in order to isolate the parameter of interest ($d$), we now  have to account for the amplification effect. Our signal is boosted from $d\, \la {\bf A} \ra $ to $d\, A_w$ and the correlation matrix changes to $C'$, and is now a smaller approximately $ \gamma N \times \gamma N$ matrix. The Fisher information is given by \cite{jordan}
\be
{\cal I}_{wv}  =  A_w^2 \sum_{i,j} [C'^{-1}]_{i,j}.
\label{fiwv}
\ee
We can now treat this case like we did in the previous section. We have a new covariance matrix $C'$ with which we can make a similar decomposition, $C' = O' D' {O'}^T$.  The dimension is reduced by a factor of $ \approx \gamma$.  We make exactly the same treatment as before, calling $\sigma'^2_j$ the eigenvalues of $C'$, and $w'_j$ the new weights.

The Fisher information is now given by
\be
{\cal I}_{wv}  = A_w^2 (\gamma N) \sum_k^{~ \gamma N} \sigma'^{-2}_k w'_k,
\ee
where the weights $w'_k$ are normalized.
In order to account for the effect of the weak-value amplification, we will later give a detailed model for the precise form the weak-values take on.  For the moment, we estimate $A_w^2 = {\la \bf A \ra }^2/\gamma$, as is true in many weak-value implementation experiments.  If that is so, we have
\be
{\cal I}_{wv}  =  N {\la \bf A \ra }^2 \sum_k \sigma'^{-2}_k w'_k.
\ee
We note that the amplification factor has canceled the factor of $\gamma$ which arose due to the reduced size of the new covariance matrix. Comparing this relation to (\ref{FI}), accounting for the multiplication of the Fisher information with ${\la \bf A \ra }^2$, we see that both scale as $N$, and the main change to the Fisher information is how the correlations are affected by the post-selection. If the randomly postselected events have the same type of correlations as the non-post-selected case, then the Fisher information is comparable - it is still a weighted average of (a smaller number of) inverse eigenvalues.  If the correlations are reduced because retained measurements are further separated in time reducing temporal correlations, for example, then the Fisher Information could be larger and WVA would possess an advantage over the direct measurement approach. We will soon see that this is indeed the case. 

\section{Optimal Partitioning Measurement}
\label{improved}
We next consider improving on the weak-value amplification scheme by incorporating the discarded measurement results in our estimation strategy. This involves optimally implementing a partitioning measurement so that all output channels and resulting correlations are used.  We will refer to this as an optimal partitioning class measurement (OPM), which we can compare to the simpler WVA case.

For an $M$ dimensional system, there are $M$ possible outcomes of the second projective measurement on the system, $|f_1\ra, |f_2\ra, \ldots, |f_M\ra$.  For each of those possibilities, there is a weak-value, so $A_{w,f_j}$, $j = 1, \ldots, M$.   The distribution of events is assumed to be a multi-variable Gaussian distribution $P(\{ s_j\}|d)$, with mean ${\vec \mu} = {\vec A}_w d$ and covariance matrix $C$.  Here, we define a vector of weak-values, ${\vec A}_w$, associated with each outcome with elements $A_{w,f_j}$, where $j=1, \ldots M$. 
We will now focus on the $M=2$ case.

Given $N$ measurement outcomes, the selection tags $\gamma N$ of the outcomes with one post-selection associated with the final state $|f\ra$, and the remaining  $(1-\gamma) N$ outcomes with the post-selection  associated with the final state $| f_{\perp} \ra$.  We reorder the outcomes and label them $i = 1, \ldots \gamma N; \gamma N+1, \ldots N$.  This will not typically be the temporal ordering.  However, the first selection is associated with the weak-value $A_w$, and the second with $A_w^\perp$.  We write the covariance matrix in $2\times 2$ block form (this can be generalized for multiple partitionings in the higher dimensional case).  The probability distribution is a multi-variable Gaussian distribution.  It has mean ${\bf \mu} = (A_w d\ {\bf 1}_{\gamma N}, A_w^\perp d\ {\bf 1}_{(1-\gamma) N})^T$, where ${\bf 1}_{x}$ is a vector of $x$ 1s.  The covariance matrix in block form is 
\be
C = \begin{pmatrix} C_{11} & C_{12} \\ C_{21} & C_{22} \end{pmatrix}, \label{blochmatrix}
\ee
where $C_{11}$ (of dimension $\gamma N \times \gamma N$), and $C_{22}$ (of dimension $(1-\gamma) N \times (1- \gamma) N$) are the covariance matrices associated with the two selections, and $C_{12} = C_{21}^T$ is the correlation matrix between the selected outcomes, which has dimension $\gamma N \times (1-\gamma) N$ or $(1-\gamma) N \times \gamma N$.

The Fisher information for such a situation is given by \cite{kay2010fundamentals}
\begin{eqnarray}
{\cal I} &=&  \partial_d {\mu}^T \cdot C^{-1} \cdot \partial_d {\mu} \\
&=& {\cal I}_1 + {\cal I}_2 + {\cal I}_3\\
&=& A_w^2 \sum_{ij} ([C^{-1}]_{11})_{ij} +
(A_w^\perp)^2 \sum_{ij} ([C^{-1}]_{22})_{ij} \label{correlatedfi}  \\
&+&  A_w A_w^\perp \left( \sum_{ij} \nonumber ([C^{-1}]_{12})_{ij} + \sum_{i'j'} ([C^{-1}]_{21})_{i'j'}\right),
\end{eqnarray}
where the sums run over the appropriate ranges.
It is important to note that $[C^{-1}]_{kl}$, where $k,l = 1,2$ refers to the block-matrix form of the inverse of the entire $C$ matrix, not the inverses of each of the sub-blocks. Just as in the simple example of the $2 \times 2$ covariance matrix, the total information exceeds using either selection state alone, however, there may be cases where all the information is in one of the selection states, and the other may be discarded without any loss in variance.  We will see such an example in the next section.

It is also of interest to find the optimal estimators in this case, using maximum likelihood methods.
 Each event is filtered according to the selection state, and is associated with a mean of that weak-value times $d$.  The covariance matrix is assumed to have the same values as before (with a mean meter shift of $A_w d$ or $A_w^{\perp} d$) in the weak measurement limit, but the indices are relabeled to put the matrix into block form, associated with each weak-value.

We can find the optimal estimator by solving for  the value of $d$ that maximizes the log-likelihood, $\log P(\{ s_j\}|d) = -({\vec s} - {\vec A}_w d)^T \cdot C^{-1}\cdot ({\vec s} - {\vec A}_w d) + const$ with respect to the parameter $d$ \cite{jordan}.  The maximum likelihood estimator is given by
\be
{\hat d} = \frac{{\vec A}_w^T \cdot C^{-1} \cdot {\vec s}}{{\vec A}_w^T \cdot C^{-1} \cdot {\vec A}_w},\ee
where ${\vec s}$ is a vector of outcomes.

For the special case of 2 selection states, with weak-values $A_w$, and $A_w^\perp$, so ${\vec A_w} = (A_w \, {\bf 1}_{\gamma N}, A_w^\perp \, {\bf 1}_{(1-\gamma) N})$, we assume that each selection has outcomes $1, \ldots \gamma N$, and $\gamma N+1, \ldots N$.   The inverse covariance matrix takes the form of a $2 \times 2$ block matrix.  We further break the vector of outcomes in two, corresponding to each selection label, ${\vec s} = ({\vec s_1}, {\vec s_2})$.  To give an explicit expression, we express the estimator as a ratio, ${\hat d} = {\cal N}/{\cal D}$, in which the numerator ${\cal N}$ is given by 
\begin{widetext}
\begin{eqnarray}
{\cal N} &=& (A_w \, {\bf 1}_{\gamma N}, A_w^\perp \, {\bf 1}_{(1-\gamma) N}) \cdot  \begin{pmatrix} [C^{-1}]_{11} & [C^{-1}]_{12} \\ [C^{-1}]_{21} & [C^{-1}]_{22} \end{pmatrix} \cdot \begin{pmatrix} s_1 \\ s_2 \end{pmatrix}, \label{oestimators}\\
&=& A_w \sum_{i = 1}^{\gamma N} \left( \sum_{j = 1}^{\gamma N}  ([C^{-1}]_{11})_{ij} s_{1,j} + \sum_{j = \gamma N+1}^{ N}  ([C^{-1}]_{12})_{ij} s_{2,j}\right)\\
&+& A_w^\perp \sum_{i = \gamma N+1}^N \left( \sum_{j = 1}^{\gamma N}  ([C^{-1}]_{21})_{ij} s_{1,j} + \sum_{j = \gamma N+1}^{ N}  ([C^{-1}]_{22})_{ij} s_{2,j} \right) \nonumber,
\end{eqnarray}
\end{widetext}

and the denominator $\cal D$ is the Fisher information, (\ref{correlatedfi}).

\section{Exactly solvable model}
\label{solvable}
\subsection{Direct case}
We now consider a simplified model for correlated noise, in which the covariance matrix corresponds to a combination of uncorrelated noise in time ($a$) and perfectly correlated noise ($c$).  The $N \times N$ covariance matrix 
$C$ is then a sum of a diagonal matrix and a matrix of identical elements, 
\be
C_{ij} = a \delta_{ij} + c,
\label{cmatrix}
\ee
where the constant term in equation 34 reflects a shift common to all elements of the data set (e.g., a systematic error), which is drawn from a zero-mean gaussian distribution with a variance of c.

Equation \ref{cmatrix} has a simple eigensystem, because the characteristic equation for the eigenvalues, ${\rm det}( C - \sigma^2 I)=0$ may be solved by substituting $\sigma^2 = c \sigma'^2 + a$, and noting that $\sigma'^2$ solves for the eigenvalues of the matrix of all 1s.  The latter matrix has one eigenvalue of $N$, and the rest 0.  The first normalized eigenvector is ${\bf e}_1 = (1, 1, \ldots, 1)/\sqrt{N}$, corresponding to the $N$ eigenvalue, and the other $N-1$ eigenvectors corresponding to the 0 eigenvalue are constructed orthogonal to ${\bf e}_1$.

This result indicates that the eigenvalues of $C$ are
\be
\sigma^2_j  = (N c + a, a, a, \ldots, a),
\ee
with the same eigenvectors mentioned above.  The positive semidefinite condition requires that $c \ge -a/N$. We can now apply the procedure outlined above by writing the orthogonal matrix formed from the orthonormal eigenvectors as
\be
O = ({\bf e}_1, {\bf e}_2, \ldots, {\bf e}_N).
\ee
The vector $v_k = \sum_i O_{i,k} = \sum_i ({\bf e}_k)_i$ may be computed for this model by using the orthonormality conditions of the vectors ${\bf e}_j$.  
This is because the sum runs down the column of each unit basis vector.  Using the properties
\be
\sum_i ({\bf e}_1)_i = \sqrt{N}, \quad {\bf e}_1 \cdot {\bf e}_j = \sum_i ({\bf e}_j)_i =0,
\ee
for every unit vector with $j \ne 1$, we find
\be
{\bf v} = (\sqrt{N}, 0, 0, \ldots, 0), \quad {\bf w}= (1, 0, 0, \ldots 0),
\ee
for this model.   Noting the Eq.~(\ref{normalization}) is correct, we find the Fisher information (\ref{FI}) with the shifted mean is
\be
{\cal I}_c = {\la \bf A \ra}^2 \frac{N}{N c + a}.
\label{Isolve}
\ee
We can compare this to the case of uncorrelated noise, in which
\be
{\tilde C}_{ij} = (a + c) \delta_{ij},
\ee
and where the Fisher Information is
\be
{\cal I}_{uc} = {\la \bf A \ra}^2 \frac{N}{c + a}.
\label{Isolve2}
\ee
We see the effect of the correlations is to contribute to the denominator such that in the limit of $N \rightarrow \infty$ the information (\ref{Isolve}) saturates at $1/c$. 

 It is interesting to compare this result with the equal weighting estimator, Eq.~(\ref{direct}).  Since there is only one non-zero weight, the result is
\be
{\la \bf A \ra}^2 {\rm Var}[{\hat d}] = \frac{1}{ N^2} \sum_{ij}( a \delta_{ij} + c) = \frac{a}{N} + c = {\la \bf A \ra}^2 {\cal I}^{-1}.
\ee
We see in this case, the improved estimation strategy does not help reduce the variance beyond simply averaging the data. This result is easily predicted from the structure of the covariance matrix as previously shown. As $c \rightarrow -a/N$, the information diverges, or the variance vanishes.  On the other hand if the correlations are positive ($c>0$), the information is invariably reduced. This reduction is easily understood as being the result of an unknown offset ($c$) which is not reduced by increasing the number of measurements ($N$) (in the limit of $N \rightarrow \infty$, the variance limits to $c$ rather than zero, see also Ref.~\cite{pang2016protecting}).

\subsection{Weak-Value case}
We can apply the same model as above to the Fisher information in the weak-value case. In this case,  along with the measurement of the parameter of interest, a second partitioning measurement is performed, which divides the first set of measurements into a retained and a discarded partition. As we saw before $d  {\la \bf A \ra } \rightarrow A_w  d$.  However, with this particular noise model, the covariance matrix is exactly the same as in the non-post-selected case (because every event has correlation $c$ with every other event, and auto-correlation $a$).  Consequently, we can just apply the above results to Eq.~(\ref{fiwv}) to find the post-selected Fisher information to be
\be
{\cal I}_{wv} = A_w^2 \frac{\gamma N}{a + \gamma N c}.
\label{iwv}
\ee
Replacing $A_w^2 = {\la \bf A \ra }^2/\gamma$, the post-selection probability cancels in the numerator, and we find the same Fisher information (\ref{Isolve}), but with the effective change of 
\be
c \rightarrow \gamma c.
\label{cscale}
\ee  
That is, \textit{post-selection reduces the size of the correlation}.  However, the resulting advantage depends on the kind of noise:  If $c <0$, the post-selection increases the variance, whereas if $c >0$, the post-selection reduces the variance of the optimal estimator.  The later is in accordance with the findings of Ref.~\cite{feizpour2011amplifying} for this model.  We note that, as before, there is no difference between the optimal estimator and the equal-weighting estimator.  

We note that while $\gamma$ can be made arbitrarily small, the Fisher information is bounded by the necessity of sampling some high-information content events, selected from the covariance matrix elements $a+c$, giving the bound
\be
\mathcal{I}_{wv}  \le \la {\bf A}\ra^2 \frac{N}{a+c}.
\ee
This result represents an enormous suppression of the detrimental effects of 
noise accompanied by a significant reduction in the size of the data set. The WVA technique is able to recover the performance of the conventional method in the uncorrelated noise limit by simply selecting a small enough $\gamma$. If for some experimental reason, $\gamma$ is bounded to be larger than some minimum, $\gamma_{min}$, gives a practical limit to the noise reduction (\ref{cscale}).

\subsection{Using the other selection - optimal partitioning measurement}

We now apply the results of Sec.~\ref{improved} to our model to see how much the Fisher information may be improved by the optimal partitioning method, which involves optimally weighting both the ``post-selected" and the ``post-selection rejected" partitions of the dataset in order to estimate the parameter of interest. Comparing this approach to the WVA strategy will tell us how much information was discarded by the post-selection step in WVA. The Fisher information may be explicitly evaluated in our exactly solvable model.   This is because the exact inverse of matrix (\ref{cmatrix}) is given by 
\be
C^{-1}_{ij} = \frac{(a + c N) \delta_{ij} - c}{a^2 + N a c},
\label{optimal_result}
\ee
as can be checked by direct calculation, $C C^{-1} = I$.  It is straightforward to see that summing over both indices in (\ref{optimal_result}) returns the Fisher information (\ref{Isolve}).   The Fisher information with weak-value amplification, 
Eq.~(\ref{correlatedfi}) in this special case, has the exact form:
\be
{\cal I} = \frac{a N [ \gamma A_w^2 + (1-\gamma) (A_w^\perp)^2] + c \gamma (1-\gamma) N^2 (A_w - A_w^\perp)^2}{a^2 + N a c }.
\label{fidecomp}
\ee

\begin{figure}[H]
\includegraphics[width=\columnwidth]{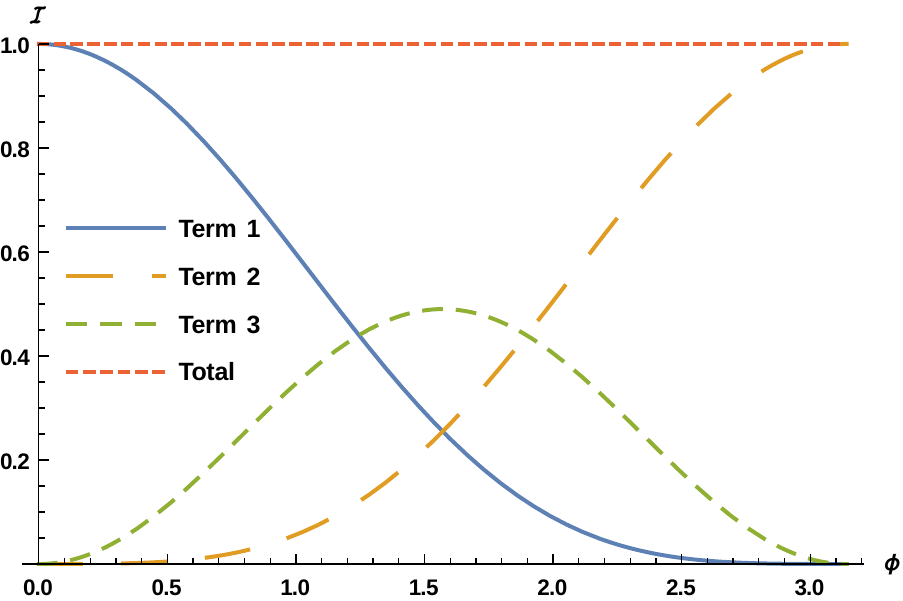}
\caption{Different contributions to the Fisher information in Eq.~(\ref{correlatedfi}) versus $\phi$, in units of $N/a$, the uncorrelated Fisher information.  Term 1 (${\cal I}_1$) is proportional to $A_w^2$; term 2 (${\cal I}_2$) is proportional to $(A_w^\perp)^2$, and term three (${\cal I}_3$) is proportional to $A_w A_w^\perp$.  We take $N=100$, $c/a = 0.5$.
}
\label{fisherdecomposefig}
\end{figure}

In order to compare the different methods, 
we adopt the standard model of the weak-value, taken from \cite{Howell}:  
\be
A_w = - \cot(\phi/2),\ A_w^\perp = \tan(\phi/2),\ \gamma = \sin^2(\phi/2),
\label{spinmodel}
\ee
where $\phi/2$ is the overlap angle between $|f\ra$ and $|i\ra$, the pre- and post-selected states.  This model leads to the simplification of the three terms defined in (\ref{correlatedfi}) as
\begin{eqnarray}
{\cal I}_1 &=& \frac{a N \cos^2(\phi/2) + c N^2 \cos^4(\phi/2)}{a^2 + N ac}, \\
{\cal I}_2 &=& \frac{a N \sin^2(\phi/2) + c N^2 \sin^4(\phi/2)}{a^2 + N ac}, \\ 
{\cal I}_3 &=& \frac{2 c N^2 \sin^2(\phi/2)\cos^2(\phi/2)}{a^2 + N ac}. 
\end{eqnarray}
The sum of these three terms is
\be
{\cal I}_s = \frac{N}{a}.
\label{fitotall}
\ee
Remarkably, the total Fisher information is now independent of the value of $\phi$, and is larger than the uncorrelated Fisher information.  Assuming $c >0$, we recall that just using the one selection state had the effect of reducing $c \rightarrow \gamma c$. We see here that adding in the other selection state and their correlation allows us to eliminate the effect of $c$ entirely. In Fig. \ref{fisherdecomposefig}, we show the various contributions to the combined Fisher information using both output selections.   The Fisher information (\ref{correlatedfi}) comes from the three terms from each sub-block. We note that if it is the case that $c<0$, then it is clearly advantageous to use the direct measurement scheme. Below we tabulate the Fisher Information of the three approaches we consider in white and slow noise limits.

\begin{center}
 \begin{tabular}{|c c c||} 
 \hline
 Approach | & Uncorrelated Noise | & Correlated Noise \\ [0.5ex] 
 \hline\hline
 FI Direct & $\frac{N}{a+c}$ & $ \frac{N}{a+Nc}$ \\ 
 \hline
 FI WVA & $\frac{N}{a+c}$ & $\frac{N}{a+c}$  \\
 \hline
 FI OPM & $\frac{N}{a+c}$ & $\frac{N}{a}$  \\ [1ex] 
 \hline
\end{tabular}\par
\bigskip
\label{tab:table-name}
Table 1: In this table, the Fisher Information is summarized in the uncorrelated and correlated noise limits for the three different strategies we consider, the direct approach, the weak value approach, and the optimal partitioning approach. $N$ is the number of measurements carried out, ($\gamma N$ is the number of retained measurements in the WVA approach), $a$ is the variance of the white noise, and $c$ is the variance of the unknown systematic. OPM removes all effect of $c$, while WVA reduces this effect by a factor of $N$, recovering exactly the variance achievable in the absence of correlations. 
\end{center}

Summarizing the chart above, we have found that all three approaches are identical in terms of measurement performance in the uncorrelated noise limit. Whichever is experimentally easier to implement, therefore, has the advantage. In the correlated noise limit, WVA has a clear advantage over the direct approach, confirming the work of Feizpour, Xing, and Steinberg \cite{feizpour2011amplifying}. What Feizpour \textit{et. al.}  did not consider, however, is that unlike in the white noise limit, once there are correlations between measurements, retaining all partitions and the correlations between them becomes advantageous.  In fact, in the same regime where WVA has a real quantitative advantage over the conventional approach, it turns out to be slightly inferior to the optimal partitioning method.

\subsection{Limiting optimal estimators}
We see from the above analysis that with the OPM approach, the selection angle $\phi$ simply changes how the information is distributed in the various outcomes.
It is instructive to focus on the form of the optimal estimators (\ref{oestimators}) in the two extreme limits for our exactly solvable model:  $\gamma \ll 1$, the weak-value amplification limit, and $\gamma = 1/2$, the balanced limit, in order to understand how the optimal partitioning approach is able to completely eliminate $c$. 

In the very unbalanced limit, the term proportional to $A_w$ dominates the Fisher information.  The weak-values are given approximately by $A_w \approx -2/\phi$, $A_W^\perp \approx \phi/2$, and the asymmetry is given by $\gamma \approx \phi^2/4$.
We find the optimal estimator to be
\be
{\hat d} \approx {\hat d}_{wv} - \frac{A_w c \gamma}{a+N c} \left(\sum_{j=1}^{\gamma N} s_{1,j} + \sum_{j=\gamma N+1}^{N} s_{2,j} \right),
\ee
where the {\it weak-value estimator} ${\hat d}_{wv}$ is defined as
\be
{\hat d}_{wv} = \frac{A_w}{N} \sum_{j=1}^{\gamma N} s_{1,j}.
\ee
The total (correlated) optimal estimator is just the weak-value estimator, plus another term involving the sum of all the data, whose average is approximately 0, and its prefactor vanishes as $c \rightarrow 0$.  The additional term is able to account for the (known) correlations in the system and make a further suppression of the variance, at the cost of having to process all collected data.

We now turn to the balanced case, where $\phi = \pi/4, \gamma = 1/2$, and the weak-values are $A_w = -1$, $A_w^\perp = 1$.  In this case, the weighting prefactors in front of the collected data cancel out, and we find the estimator
\be
{\hat d}_{b} = \frac{1}{N}  \left(\sum_{j=1}^{N/2} s_{2,j} - \sum_{j=N/2+1}^{N} s_{1,j} \right), \label{subtract}
\ee
that is, we simply subtract the data from output channel 1 from that of output channel 2, and divide by $N$. This result (\ref{subtract}) is identical to the ``background subtraction'' technique commonly used in experimental labs to eliminate correlated noise.  We discuss this in more detail in the next section.

\section{Background Subtraction}
Background subtraction typically involves partitioning measurement outcomes into two types: measurements of signal plus background noise, and measurements of just background noise.  By comparing the two partitions, unknown offsets or errors in calibration can be corrected. Background subtraction can also be used to suppress temporally correlated noise by sampling the noise background at least once per correlation time and subtracting off the slowly evolving offset.  Unfortunately, optimizing how often to sample the background noise is usually a hard problem. This is because usually both white and slow noise are present, and oversampling the background (sampling more than once per correlation time) ceases to reduce either. Optimizing the amount of background subtraction (as opposed to normal measurements) is only possible with knowledge of the correlation time of the noise. This makes it preferable, when possible, (as it is in our example), to alternate the sign of the signal instead of chopping it on and off. In this variant of background-subtraction, measurement outcomes are partitioned into two types: measurements of a background noise plus signal, and measurements of background noise minus signal. By subtracting the two partitions and dividing by two (just like in section \ref{2dexample}), we can eliminate any slow noise present while also averaging down the white noise. 

While we have considered partitioning measurements using post-selection, it is possible to achieve the same result by alternating the preparation. It is worth noting that the additional information gained by this more complex use of the partitioning measurement is very marginal compared to WVA. Furthermore, we believe there may be situations where alternating the preparation is experimentally challenging. For example, in certain optical implementations of WVA, where different degrees of freedom of the same photon can be used to encode signal information and post-selection information, background subtraction can occur at much higher rates than signal preparation \cite{Dixon2009}.

\section{Numerical Investigation}
\label{numerics}
We will now retun to our original (and more experimentally relevant) noise model. This noise model represents a familiar scenario in a laboratory. Measurements occur sequentially in time and are compiled into a data set $\{s_i\}$. As before, a measurement outcome $s_i$ can be decomposed into the parameter of interest, $d$, and a Gaussian distributed, zero-mean, random variable, $x_i$, such that
\be
s_i = d + x_i,
\ee
where $i = 1,2, ..., N$. The covariance matrix is
\be
C_{i,j} = a\delta_{i,j} + c e^{-|i-j|\Delta t / \tau}.
\end{equation}
Here, $\Delta t$ sets the time between subsequent measurements, $\tau$ represents the correlation time of the noise, and the ratio of $a$ to $c$ sets the relative amount of white noise to slow noise (except in the limit where $\tau \rightarrow 0$ where all noise is white). For simplicity we will use the unitless quantity $\eta =\tau / \Delta t$ (the ``average'' number of correlated measurements) to represent the correlation time of the noise. 

This noise model captures an experimentally common scenario, wherein there is a white noise floor (represented by $a$) and some correlated noise arising due to experimental imperfections (represented by $c$). Overcoming technical noise of this kind is quite challenging, hence the interest in a technique that is robust against it. 

\begin{figure}[tbp]
\includegraphics[width=\columnwidth]{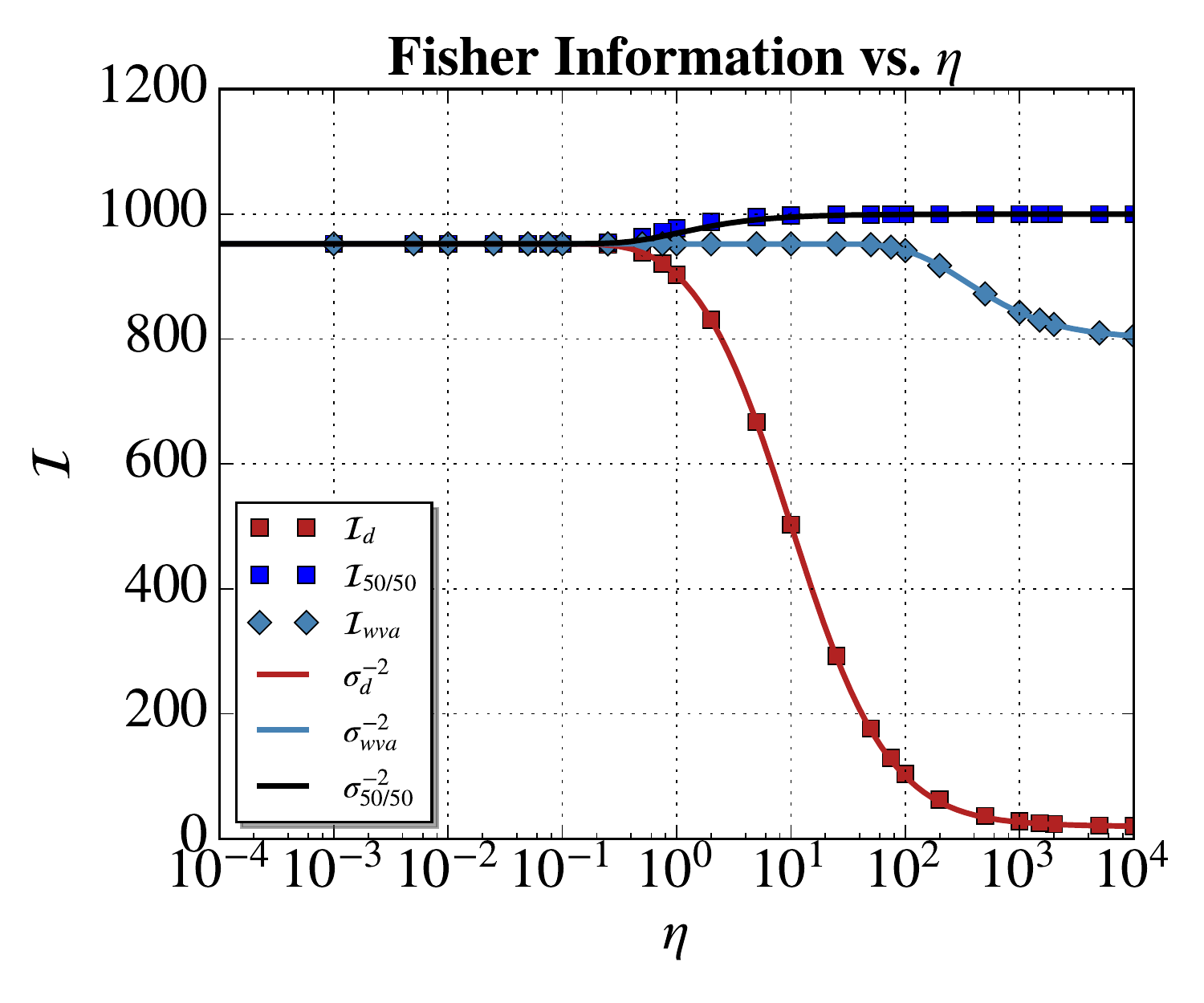}
\caption{The Fisher information for the direct method ${\cal I}_d$, the partitioning WVA method ${\cal I}_{wva}$, and the partitioning balanced method (corresponding to background subtraction) ${\cal I}_{50/50}$ are plotted for different values of dimensionless correlation time $\eta = \tau/\Delta t$. The inverse variance of the equal-weighting estimator is also plotted for the three approaches considered. In this figure $N=1000, a=1, c=0.05$, and $\gamma=.005$. The knee in the performance of the WVA approach occurs at $\eta = 1/\gamma$, where retained measurement outcomes begin to become correlated again. }
\label{josiah_figure}
\end{figure}

We have already considered two limiting cases of this generalized noise model. In the ``white noise limit," $\eta \rightarrow 0$ and measurement outcomes are uncorrelated. In the opposite limit, $\eta \rightarrow \infty$ and the covariance matrix reduces to the directly solvable one from section \ref{solvable},
\be
C_{i,j} = a \delta_{i,j} + c.
\ee
We refer to this as the ``slow noise limit." This limit represents taking the correlation time of the noise to infinity resulting in a scenario where estimation error is increased by some unknown offset, $c$. We have previously treated the two limits of our noise model for all three measurement approaches considered (direct, WVA, and OPM). We expect the effects of the correlations to grow as the correlation time (or $\eta$) increases, smoothly connecting our two limits for all the measurement approaches considered. Finding the Fisher Information in this intermediate regime, however, is challenging due to the complex structure of the covariance matrix. In order to find the Fisher Information, we generate large covariance matrices with the appropriate structure and invert and sum them while varying the correlation time. This allows us to smoothly connect the two limits. In figure \ref{josiah_figure}, the Fisher Information for both the WVA method, the direct method, and background subtraction (equivalent to OPM) is plotted as a function of the average number of correlated measurements, $\eta$. In each scenario, the Fisher Information is found through construction of the covariance matrix that would be produced by the measurement design and then inverted and summed. For comparison, the inverse variance of the mean associated with the equal-weighting estimator is also found analytically and plotted in solid lines. In the two limits ($\eta \rightarrow 0$ and $\eta \rightarrow \infty$) the two estimators are equivalent. In the intermediate regime, the equal-weighting estimator is no longer strictly optimal. It deviates only very slightly from the Fisher Information as a result of the finite size of the covariance matrices we consider (due to the boundaries of our data set certain measurements are correlated to more measurements than others).  In figure \ref{josiah_figure}, as expected, as $\eta \rightarrow 1$, the Fisher information in the direct method degrades swiftly, while the Fisher Information of the weak-value method is unaffected. Only after $\eta $ approaches $ 1/\gamma$, ($\gamma$ was arbitrarily chosen to be 1\%), does the performance of the weak-value method begin to diminish. If $\gamma$ is fixed, then in the limit of $\eta \rightarrow \infty$ the minimum variance becomes $a/N + \gamma c$. Even in the intermediate regime, with weak correlations, post-selection suppresses the detrimental effects of the correlations by increasing the time between retained measurement outcomes. 

These results confirm the work of Feizpour, Xing, and Steinberg who argued that WVA-based measurement estimation strategies afforded an advantage in a correlated noise limit over direct methods. They also agree with Ferrie and Combe's observation \cite{ferrie2014weak} that WVA does not always achieve globally optimal performance. We see that the OPM/BS technique slightly outperforms WVA as soon as the correlation time becomes comparable to the time between measurements. This difference can be substantial if the minimum probability of post-selection is large; however, if it is on the order of $\frac{1}{\gamma}$ it is quite small. In figure \ref{josiah_figure} the relative size of $a$ and $c$ that represent the white and slow portions of the noise was chosen in order to visually differentiate the different approaches. In typical situations, $c$ is much less than $a$, for otherwise an experimentalist would be easily able to detect it in preparation for the experiment. If, however, the value of $c$ is on the order of $a/N$, then an experimentalist would have to spend more time characterizing the noise then performing the experiment - an unlikely scenario in all but the most heroic metrology experiments. A realistic value for $c$ is therefore on the order of $a/N$, and we see that in this scenario the additional advantage afforded by optimally using all the data is
\be
\Delta I = {\frac{N}{a}}-{\frac{N}{a+c}}\approx {\frac{1}{a+a/N}}.
\ee
This cost in precision limits to $1/a$ as $N\rightarrow \infty$ and is equal to the information which would be gained by a single additional measurement in this scenario. These numerical results confirm that the conclusions reached using our simplified model for correlated noise hold generically in more realistic laboratory situations with noise environments with time decaying correlations. 

\section{Conclusions}
\label{conclusions}
We have investigated how parameter estimation is affected by the introduction of a second partitioning measurement.  Using a realistic model for additive, Gaussian, time-correlated, noise we found that, for all the cases considered, introducing a second partitioning measurement affords an advantage in the Fisher Information over a direct measurement method (no second partitioning measurement) once the correlation time of the noise becomes longer than the measurement rate. Furthermore, we have found that if one or more output channels from the partitioning measurement are filtered (corresponding to a WVA-like post-selection) the majority of this advantage is retained. Thus, WVA can help dramatically suppress the detrimental effects of slow noise, recovering the performance achieved in the white-noise limit by effectively decreasing the correlations between retained measurement outcomes. The informational cost of the post-selection step is studied by comparing the Fisher Information of the WVA approach with an optimal partitioning measurement scheme which utilizes all the data and correlations between partitions. The cost of discarding all but one partition is found to be related to the ratio of the magnitudes of the slow and white noise. We argue that in a typical laboratory situation, this cost will be comparable to the information gained by performing a single extra measurement and is therefore negligible. An analysis of the estimators used by OPM leads us to the realization that in the balanced case OPM is equivalent to probabilistic background subtraction. This insight provides a unified framework for understanding when partitioning-class measurements, and specifically weak value amplification, can be useful. In conclusion, in experimental settings with time-correlated noise, if background subtraction or other optimal partitioning measurement methods are technically challenging to implement, or if a substantially reduced data-set is desirable, WVA vastly outperforms conventional measurement approaches and is near-optimal.

\acknowledgements
This work was supported by the Army Research Office Grant
No. W911NF-13-1-0402, DRS Technologies, Chapman University, NSERC, CIFAR, Northrop Grumman Aerospace Systems \textit{NG Next}, and the Fetzer Franklin Fund of the John E. Fetzer Memorial Trust.

\bibliography{correlatednoise}

\end{document}